# Cluster Dose Prediction in Carbon Ion Therapy: Using Transfer Learning from a Pretrained Dose Prediction U-Net

Miriam Schwarze[1,2,*], Hui Khee Looe[2], Björn Poppe[2], Leo Thomas[1] and Hans Rabus[1]

[1] Physikalisch-Technische Bundesanstalt, Berlin, Germany
[2] Carl von Ossietzky Universität Oldenburg, Oldenburg, Germany

* Corresponding author, miriam.schwarze@ptb.de

## Abstract

The cluster dose concept offers an alternative to the radiobiological effectiveness (RBE)-based model for describing radiation-induced biological effects. This study examines the application of a neural network to predict cluster dose distributions, with the goal of replacing the computationally intensive simulations currently required. Cluster dose distributions are predicted using a U-Net that was initially pretrained on conventional dose distributions. Using transfer learning techniques, the decoder path is adapted for cluster dose estimation. Both the training and pretraining datasets include head and neck regions from multiple patients and carbon ion beams of varying energies and positions. Monte Carlo (MC) simulations were used to generate the ground truth cluster dose distributions. The U-Net enables cluster dose estimation for a single pencil beam within milliseconds using a graphics processing unit (GPU). The predicted cluster dose distributions deviate from the ground truth by less than 0.35%. This proof-of-principle study demonstrates the feasibility of accurately estimating cluster doses within clinically acceptable computation times using machine learning (ML). By leveraging a pretrained neural network and applying transfer learning techniques, the approach significantly reduces the need for large-scale, computationally expensive training data.

## 1. Introduction

In ion beam therapy treatment planning, the biological effects of radiation are currently quantified by radiobiological effectiveness (RBE), defined as the ratio of the dose of radiation under consideration to the photon dose required to achieve the same biological outcome. The RBE strongly depends on the linear energy transfer (LET) of the radiation, which increases for carbon ions along their path and reaches its maximum in the Bragg peak region [1]. Treatment planning in carbon ion therapy accounts for this varying LET by optimizing the RBE-weighted dose, defined as the product of RBE and the absorbed physical dose [2]. Various semi-empirical models are used to determine the RBE, including the local effect model (LEM) [3–7] in Europe, the mixed-beam model [8,9], and the microdosimetric kinetic model (MKM) [10–13] in Japan. The RBEs calculated by the models differ by up to 15% from each other, and there is currently no consensus on which model provides the most accurate representation of RBE [14–16].

Faddegon et al. [17] introduced the cluster dose as an alternative method to the RBE-weighted dose. This concept uses the correlation between the biological effects of ionizing radiation and the spatial distribution of energy deposits with ionizations in biomolecules along the particle track [18]. Due to the small size of the targets (nanometer scale), the mean energy deposited is stochastic in nature and is described by the concepts of nanodosimetry [19].

Nanodosimetric quantities are modeled using track structure (TS) simulations. These Monte Carlo (MC) simulations utilize detailed cross-sections to determine energy depositions with nanometer-scale resolution. Detailed cross-section data are only available for a small selection of materials, which is why TS simulations are typically conducted in liquid water. In addition, the detailed simulations require significant computation time, making the simulation of macroscopic volumes practically infeasible [20,21].

The main (stochastic) physical quantity in nanodosimetry is the ionization cluster size (ICS) $\nu$, defined as the number of ionizations occurring within a target volume comparable in size to the desoxyribonucleic acid (DNA). The distribution of ICS is referred to as the ionization cluster size distribution (ICSD) $P(\nu)$, while the moments of the distribution are denoted as $M_i$. In addition, the cumulative frequency of clusters with ICS greater than k

$$F_k = \sum_{\nu=k}^{\infty} P(\nu) \qquad (1)$$

is often specified [22,23].

The concept of cluster dose provides a methodology to combine these nanodosimetric quantities with the macroscopic geometries used in treatment planning. Faddegon et al. [17] define the cluster dose of a particle of class $c$ in a voxel $j$ as

$$g_j^{I_P} = \frac{1}{m_j} \sum_{c \in \mathcal{C}_j} t_j^c I_P^c. \tag{2}$$

Here, $m_j$ represents the mass of voxel $j$ and $t_j^c$ denotes the cumulative path length of particles of class $c$ within voxel $j$, which can be obtained from a condensed-history (CH) simulation of the beam in a macroscopic volume. $I_P^c$ denotes a quantity proportional to the nanodosimetric quantity (e.g. $M_i$ or $F_k$). The classes $c$ differentiate between particle type and energy.

The ICSD and its parameters conventionally pertain to a single target and a specific impact parameter of the primary particle with respect to the target. The nanodosimetric quantities used for the cluster dose formalism are determined as the proportion of a specified larger volume (a cylinder around a portion of the particle trajectory) filled with nanometric targets in which a given ICS is obtained. This approach was first introduced by Selva et al. [24] and corresponds to averaging the nanodosimetric quantities over a beam of a given size [24]. The parameters $I_P^c$ are proportional to this variant of nanodosimetric quantities, and the number of targets per path length of the primary particle is the factor of proportionality.

By scaling the nanodosimetric quantity calculated in liquid water, the differing interaction properties of other materials can be taken into account with reasonable accuracy. Consequently, the cluster dose for materials other than water can be described by [25]

$$g_j^{I_P} = \frac{1}{\rho_0 V_j} \frac{A_0}{A_j} \sum_{c \in \mathcal{C}_j} \frac{\sigma_j^c \sigma_j^{e-}}{\sigma_0^c \sigma_0^{e-}} t_j^c I_P^c. \tag{3}$$

Here, $\rho_0$ denotes the density of the TS simulation medium, and $V_j$ is the volume of voxel $j$. The quotient $A_0/A_j$ represents the ratio of the effective mass numbers of the TS simulation medium and the voxel material. $\sigma^c$ refers to the ionization cross-section for particles of class $c$ and $\sigma^{e-}$ is the average ionization cross-section for electrons.

While a database of nanodosimetric quantities can be precalculated using TS simulations, the CH simulation for determining $t_j^c$ must be repeated for each patient and beam configuration. Depending on the geometry and beam energy, this takes several hours to a full day, which is excessive for application in clinical routines. In addition, the simulation requires significant memory storage due to the dependence of the cumulative track length on voxel, energy, and particle type. As part of integrating the cluster dose concept into the treatment planning system matRad, this CH simulation was successfully replaced by a pencil beam algorithm, significantly accelerating computation [26]. However, pencil beam algorithms are based on physical simplifications, which leads to inaccuracies, particularly at material interfaces [27,28].

In recent years, significant progress has been made in applying machine learning (ML) techniques to calculate physical dose distributions. Different studies have demonstrated that ML algorithms can achieve dose calculations within seconds, with an accuracy comparable to MC simulations [29–33]. This paper investigates whether ML algorithms are also suitable for calculating cluster dose distributions. Due to the high computational cost associated with generating cluster dose distributions in patient geometries, we utilize pretrained models originally developed for physical dose prediction and apply transfer learning techniques, which reduces the amount of training data required. The spatial distributions of physical dose and cluster dose for carbon ions are generally similar in shape. However, the Bragg peak-to-plateau ratio is more pronounced in the cluster dose distribution, reflecting the higher LET in the Bragg peak region.

## 2. Materials and Methods

### 2.1 Cluster Dose Calculation Model

To predict cluster dose distributions, we employ a U-Net-based neural network that has been pretrained on dose distribution estimation. Cluster dose distributions are computed for a selection of the same patients and beam configurations used during pretraining, as well as for an additional patient case that was not included in the pretraining data. Section 2.2 provides details regarding the generation of the underlying training dataset.

The architecture of the U-Net and the corresponding pretraining procedures are described in [33]. For a better understanding of the following description, an illustration of the U-Net architecture is attached in Supplementary Figure 1. The network receives two inputs: the patient's mass density distribution and a cluster dose distribution in homogeneous water, precomputed for the same irradiation setup. The objective of the U-Net is to transform this distribution into a patient-specific cluster dose distribution. An advantage of this approach is that the model does not need to learn all of the underlying physical processes; instead, the training focuses on capturing the material dependence of these processes. Since the cluster dose distributions in water can be precomputed once for each beam configuration independently of the individual patient anatomy, this method eliminates the need for computationally intensive patient-specific computations during deployment.

Features from the mass density distribution and the cluster dose distribution in water are extracted by two separate encoder pathways. After each encoder level, the extracted features are added. All layers in the pretrained encoders – except for the BatchNorm layers – are reused without further modification. This strategy is based on the assumption that features learned during dose prediction are also relevant for predicting cluster dose distributions. The BatchNorm layers are finetuned to adapt the normalization statistics to the distribution of the new training data. The decoder and bottleneck layers – which are responsible for the actual transformation – are also updated during training and optimized with respect to the new target distribution.

The network is trained individually for each nanodosimetric quantity $I_P$ using the Adam optimizer with an initial learning rate of $10^{-4}$. The learning rate is progressively reduced during training via exponential decay with a decay rate of 0.93 (both inferred using hyperparameter optimization). As the loss function, the mean squared error (MSE) between the predicted and the real cluster dose distribution is employed. Training is terminated when the validation MSE has not improved for ten consecutive epochs.

## 2.2 Dataset

The dataset of cluster dose distributions is created using the simulation code described in [25]. The approach comprises two parts, as shown in Figure 1. First, a CH simulation of the carbon ion beam in the patient geometry to determine the cumulative track lengths $t_j^c$ in the voxels (described in Section 2.2.1). Second, a TS simulation of the carbon ions and their secondary particles at various energy levels to determine the nanodosimetric quantities $I_P^c$ (described in Section 2.2.2).

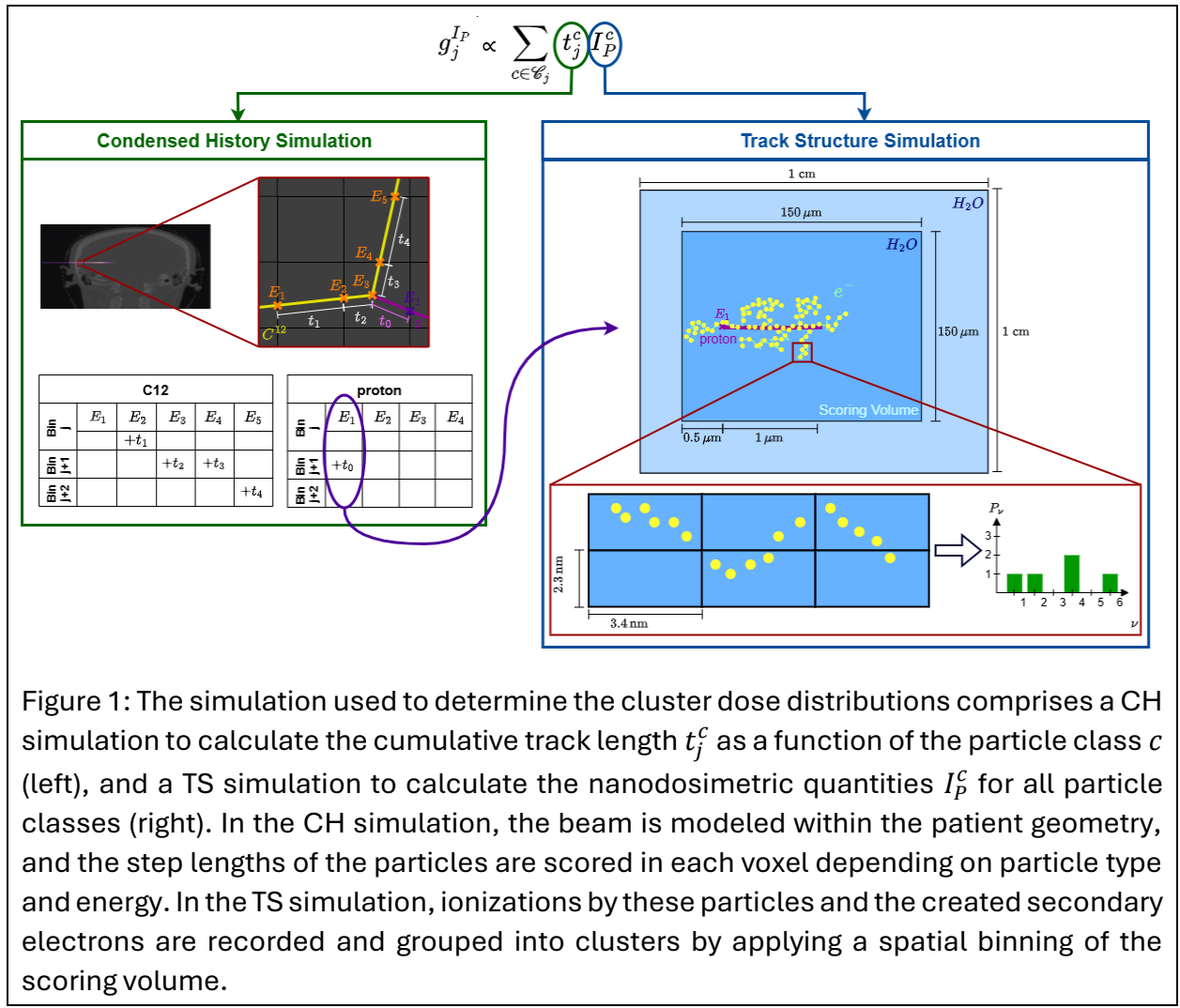

Figure 1: The simulation used to determine the cluster dose distributions comprises a CH simulation to calculate the cumulative track length $t_j^c$ as a function of the particle class $c$ (left), and a TS simulation to calculate the nanodosimetric quantities $I_P^c$ for all particle classes (right). In the CH simulation, the beam is modeled within the patient geometry, and the step lengths of the particles are scored in each voxel depending on particle type and energy. In the TS simulation, ionizations by these particles and the created secondary electrons are recorded and grouped into clusters by applying a spatial binning of the scoring volume.

### 2.2.1 Fluence Simulation

The application for calculating cumulative track lengths is based on the simulation software Geant4 [34–36]. The simulation geometry is derived from CT images of the head and neck region of various patients, available in the *GLIS-RT* dataset from the Cancer Imaging Archive [34]. For the simulation of the beam in homogeneous water, the Hounsfield unit (HU) values of all voxels are set to 0 in a copy of a Digital Imaging and Communications in Medicine (DICOM) image from the dataset (which – according to the HU scale definition – corresponds to the material water).

Using the Geant4 DICOM Reader [35], the CT images are converted into density distributions and into a voxelized geometry. Circular carbon ion beams with a diameter of 1 mm and initial beam energies between 1250 MeV and 3000 MeV are simulated. The source position is randomly selected from the source positions used in the dose dataset [33]. For each patient and beam energy, four simulations with different random source positions are performed, resulting in 300 samples.

The simulation uses the physical processes defined in the physics list of the hadron therapy example [36]. The cumulative track lengths are stored for each particle type in the form of a two-dimensional (2D) histogram. One dimension encodes the voxel index, and the other the energy bin of the particle. For each step of a particle in the geometry, the energy, the index of the voxel after the step, and the step length are determined, whereby the latter is added to the corresponding bin entry. Given that Geant4 treats a voxel boundary as a volume boundary, a step is recorded each time a voxel boundary is crossed.

The particle types and energy bins considered were determined based on a preliminary simulation of the carbon ion beam at various energies in one of the patient geometries. The particle classes used distinguish the particle types C-12, C-11, C-10, B-11, B-10, protons, alpha, He-3, H-3, and H-2, as well as 84 energy bins between 1 MeV and 3000 MeV with increasing bin width.[1] Due to the high memory requirements of the 2D histograms, the track length information is only stored within a lateral range of interest (ROI) of 16 ×16 voxels around the beam axis (for all planes in the beam direction).

To scale the nanodosimetric quantities calculated in water, the ionization cross-sections for all materials and particle classes used are determined using the functions of the *G4EmCalculator* class.

### 2.2.2 TS Simulation and Clustering

The nanodosimetric quantities are calculated using the MC TS software Geant4-DNA [37–39]. The geometry comprises water, as the detailed TS cross-sections for most particles are only available for water.

The scoring volume comprises a water cube with an edge length of 150 µm. This is embedded in a larger water cube with an edge length of 1 cm to account for potential backscattering processes. The CH cross-sections are used in this world volume, as detailed resolution is not required outside the scoring volume, which significantly reduces the simulation time. Electromagnetic processes are handled by the *G4EmStandardPhysics_option4* constructor, while the TS region utilizes the processes from the *G4EmDNAPhysics_option4* constructor. Hadronic processes are not considered here. They – along with the associated generation of secondary particles – are part of the CH simulation for determining the cumulative track lengths, whereas only the detailed spatial distribution of the particle ionizations is determined here.

---

[1] Energy bin centers: [1, 2, 3, 4, 5, 10, 15, 20, 25, 30, 35, 40, 50, 60, 70, 80, 90, 100, 125, 150, 175, 200, 225, 250, 275, 300, 325, 350, 375, 400, 425, 450, 475, 500, 525, 550, 575, 600, 625, 650, 675, 700, 725, 750, 775, 800, 825, 850, 875, 900, 925, 950, 975, 1000, 1050, 1100, 1150, 1200, 1250, 1300, 1350, 1400, 1450, 1500, 1550, 1600, 1650, 1700, 1750, 1800, 1850, 1900, 1950, 2000, 2100, 2200, 2300, 2400, 2500, 2600, 2700, 2800, 2900, 3000] MeV.

The point-like source is positioned in the scoring volume at 0.5 µm from the edge, which enables scoring the energy depositions up to 0.5 µm behind the source. The primary particle of the simulation is stopped after 1 µm, and only secondary electrons and photons are tracked thereafter. For each ionization of a particle in the scoring volume, the position of the ionization and a unique number to identify the primary particle are recorded.

The simulation is performed for all particle types considered in the fluence simulation, with the centers of the energy bins used as the initial energy. Each simulation includes 50.000 primary particles.

The ionizations are clustered using the clustering approach described by Braunroth et al. [40], in which the scoring volume is conceptionally uniformly divided into cylindrical shell segments around the beam axis, each with the same volume. The cylindrical shell segments – as in Braunroth et al. [40] – have a height of 3.4 nm and a diameter of 2.3 nm. Based on the ionizations in a simulated track, the shell segments containing ionizations are determined, and the number of ionizations per primary particle is determined in each of these shell segments, which corresponds to the ICS.

The resulting frequency distribution of targets with ionization clusters was determined per radial shell in Braunroth et al. [40] by repeating the procedure described for all simulated primary particle tracks and normalized by the number of primary particles. Here, a summation over all radial shells was applied, and the result was normalized by the path length of the primary particle, resulting in the frequency of ionization clusters per unit track length. This approach slightly differs from the one used in [41,17] in that the region used for scoring is arbitrarily large in the lateral directions so that the full radial distribution of ionization clusters is taken into account, which can extend to radial distances of several 100 nm [40].

2.2.3 Cluster Dose Calculation

The cluster dose is calculated for each voxel using Equation (3) based on the cumulative track lengths and the nanodosimetric quantities obtained in the simulations (i.e., clusters per track length). Using the voxel indices stored in the 2D histogram, the data is converted into a three-dimensional (3D) distribution.

2.2.4 Preprocessing

The 3D cluster dose and density distributions are resampled to the same voxel size of (1.332 x 1.332 x 2.5) mm using the functions of the SimpleITK library [42,43], cropped to the same ROI of (256 x 16 x 16), and normalized to the global maximum of the dataset using min-max normalization. 80 % of the samples are used for training the networks, 10 % are used as validation samples for monitoring the training process and another 10 % are retained as test samples.

## 3. Results

*3.1 Evaluation of the Cluster Dose Calculation Accuracy*

Figure 2 illustrates the distribution of the relative deviation between the real and predicted cluster dose distributions across the test dataset for the four considered nanodosimetric quantities $I_P$. The colors represent the cluster dose for the different $I_P$: blue for $g^{F_2}$, red for $g^{F_4}$, yellow for $g^{F_5}$ and green for $g^{F_7}$. All $I_P$ exhibit a similar pattern in the distribution of relative deviations, ranging from 0 % to 0.4 %.

The mean relative absolute deviations are approximately 0.1 % for all $I_P$, as summarized in Table 1. The table also lists the average root mean squared error (RMSE), which corresponds to the square root of the MSE used as the loss function. The RMSE values for all $I_P$ normalized to the maximum cluster dose value of the test dataset lie between approximately 0.003 and 0.0035.

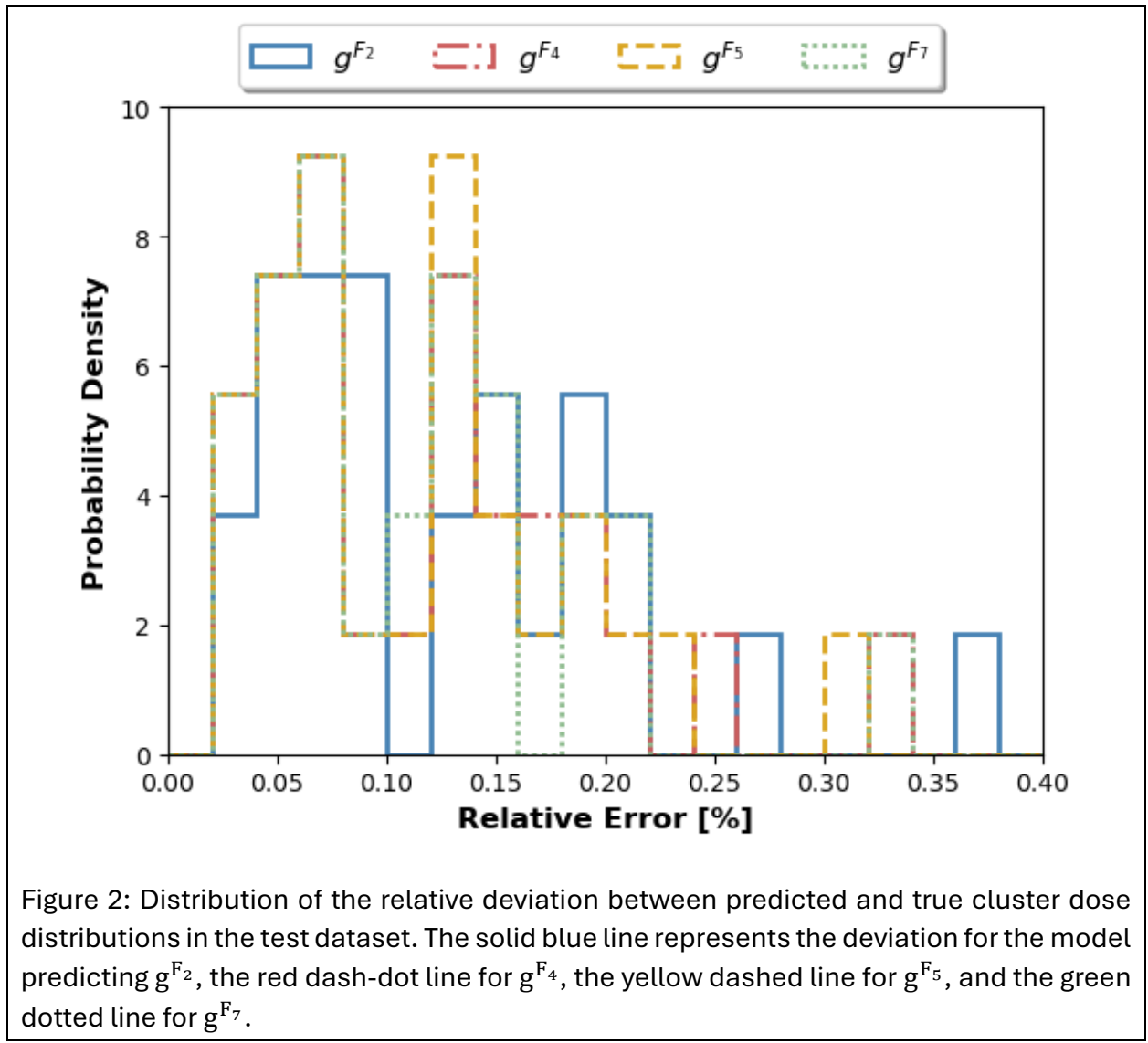

Figure 2: Distribution of the relative deviation between predicted and true cluster dose distributions in the test dataset. The solid blue line represents the deviation for the model predicting $g^{F_2}$, the red dash-dot line for $g^{F_4}$, the yellow dashed line for $g^{F_5}$, and the green dotted line for $g^{F_7}$.

Table 1: Summary of the mean voxel-wise deviations between the true and predicted cluster dose distributions in the test dataset for the five considered nanodosimetric quantities. The mean relative absolute error and the RMSE are reported, normalized to the maximum cluster dose value in the test dataset. Both the mean and standard deviation are provided for each metric.

|  | $g^{F_2}$ | $g^{F_4}$ | $g^{F_5}$ | $g^{F_7}$ |
|---|---|---|---|---|
| **Relative error [%]** | 0.12 ± 0.08 | 0.12 ± 0.07 | 0.11 ± 0.07 | 0.11 ± 0.07 |
| **RMSE [10$^{-3}$]** | 3.48 ± 1.36 | 3.17 ± 1.29 | 3.05 ± 1.24 | 3.16 ± 1.37 |

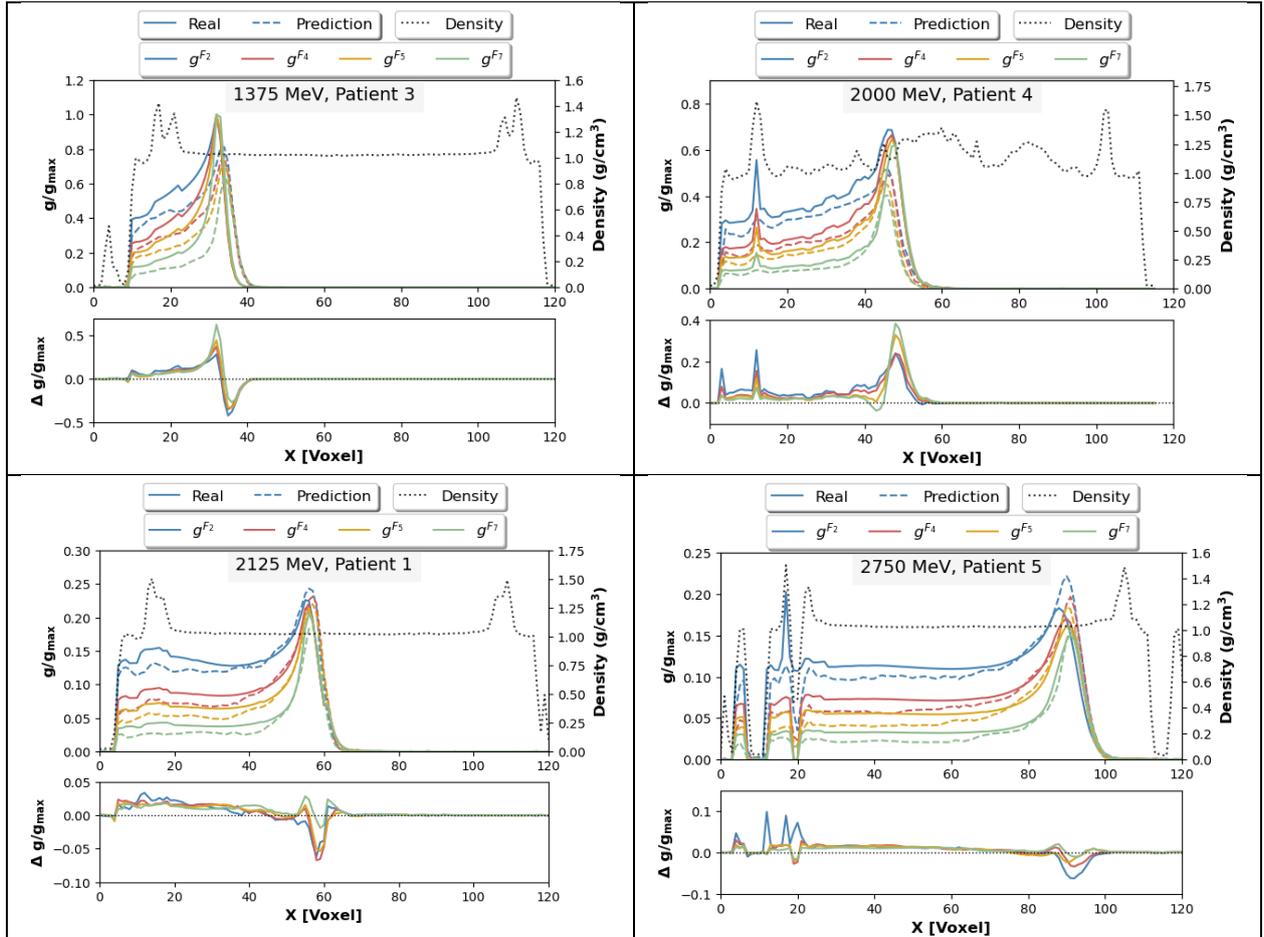

Figure 3: Comparison of the true and predicted cluster dose distributions along the beam axis for various combinations of initial beam energy and patient from the test dataset. Top: The true cluster dose profile is shown as a solid line, while the predicted profile is represented by a dashed line. All cluster dose values are normalized to the respective maximum within the test dataset. Additionally, the mass density distribution along the beam axis is shown as a dotted black line. Bottom: The difference between the true and predicted cluster dose profiles. The black dotted line at zero difference is included as a visual guide. Colors indicate the cluster dose for different nanodosimetric quantities in both panels: blue for $F_2$, red for $F_4$, yellow for $F_5$, and green for $F_7$.

Figure 3 shows four examples from the test dataset, comparing predicted and real cluster dose distributions along the beam axis for the four considered $I_P$. Each example represents a different combination of beam energy and patient, indicated at the top center of the upper panels. Cluster dose profiles are color-coded as blue for $F_2$, red for $F_4$, yellow for $F_5$, and green for $F_7$. Solid lines represent the real cluster dose profiles, while dashed lines indicate the predicted profiles in the upper panels. All cluster dose values are normalized to the maximum value of the respective test dataset. The patient's mass density distribution along the beam axis is shown as a dotted black line. In the lower panels, the deviation of the predicted from the real cluster dose distribution along the beam axis is shown using the same color coding as for the cluster dose profiles.

Accurate prediction is particularly important in the Bragg peak region, where the highest cluster dose and dose values occur. Therefore, Figure 4 provides a detailed analysis of deviations in the Bragg peak region based on four different parameters. Figure 4(a) shows deviations in the depth at which the cluster dose decreased to 90 % of the maximum value beyond the Bragg peak. Figure 4(b) illustrates deviations in the Bragg peak position perpendicular to the beam axis, while Figure 4(c) presents the relative deviation in Bragg peak height, and Figure 4(d) shows deviations in the full width at half maximum (FWHM) along the beam axis. The deviations are visualized using violin plots, where dashed lines indicate the medians and dotted lines represent the quartiles of the deviation distributions. The distributions themselves are shown as colored areas using kernel density estimates.

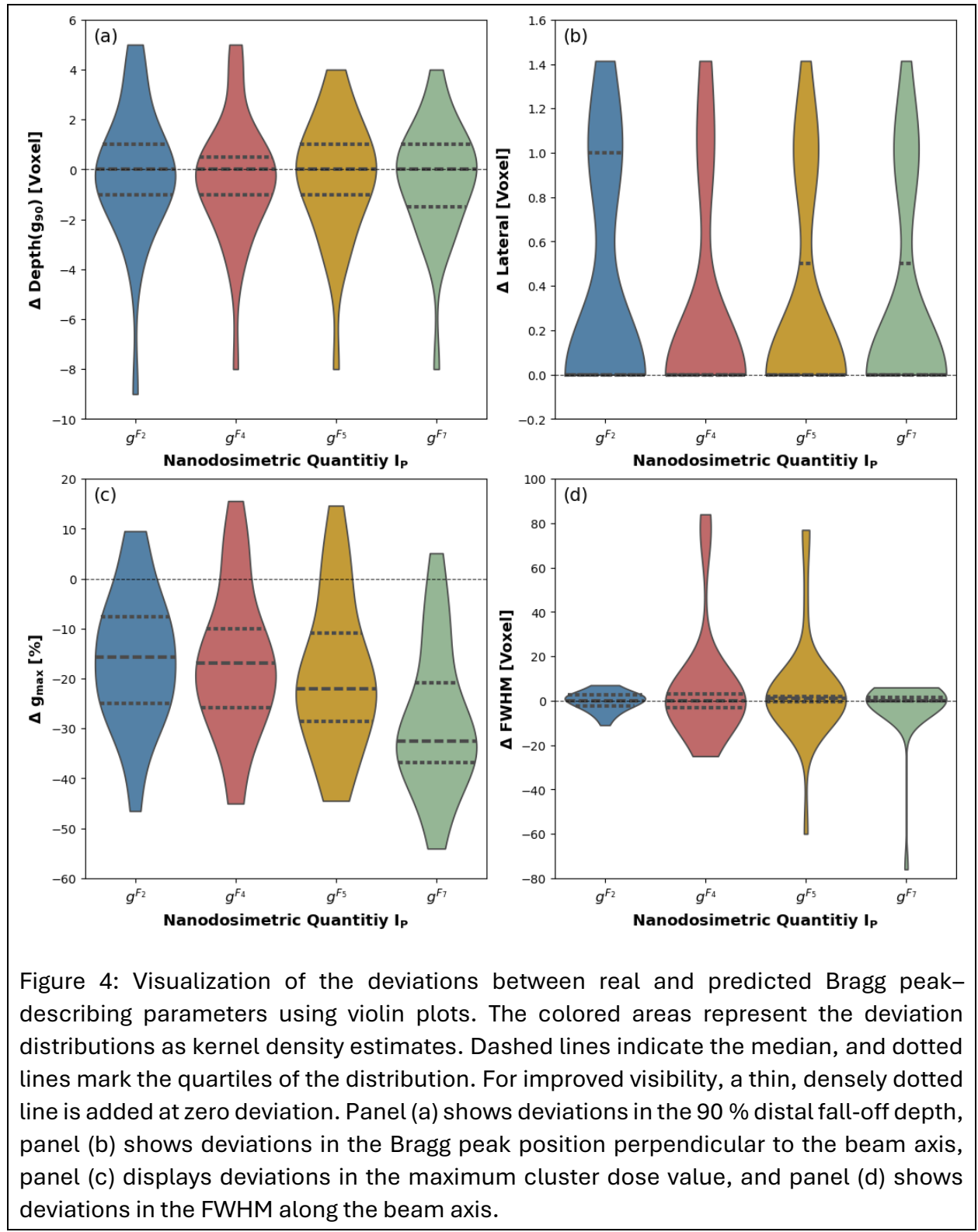

Figure 4: Visualization of the deviations between real and predicted Bragg peak–describing parameters using violin plots. The colored areas represent the deviation distributions as kernel density estimates. Dashed lines indicate the median, and dotted lines mark the quartiles of the distribution. For improved visibility, a thin, densely dotted line is added at zero deviation. Panel (a) shows deviations in the 90 % distal fall-off depth, panel (b) shows deviations in the Bragg peak position perpendicular to the beam axis, panel (c) displays deviations in the maximum cluster dose value, and panel (d) shows deviations in the FWHM along the beam axis.

*3.2 Comparison of the Cluster Dose Calculation Time*

The average computation time for generating a cluster dose distribution using the U-Net model is (11.3 ± 0.2) ms when using a GPU and (3.95 ± 0.14) s when using a central processing unit (CPU).

The training dataset was generated using Geant4 on a high-performance computing cluster, utilizing 25 CPU cores. Geant4 employs ROOT histograms for data recording in the form of histograms [44]. These histograms require significant memory during runtime, which exceeded our available resources. As a result, the simulation had to be executed four times, with each run storing histograms for only a subset of the particle types. Taking this into account, the average CPU time required to generate a single cluster dose distribution is $(7.14 \pm 2.07) \times 10^6$ s.

## 4. Discussion

The calculation of cluster dose distributions is currently a computationally intensive process. The method presented in this work – which utilizes a U-Net to determine the cluster dose distribution – significantly reduces the computational effort: When using a GPU, the calculation time for a single cluster dose distribution is only $(11.3 \pm 0.2)$ ms, compared with $(3.95 \pm 0.14)$ s when using a CPU. This corresponds to an acceleration of approximately nine orders of magnitude with GPU usage and six orders of magnitude with CPU usage compared to conventional MC simulations.

At the same time, cluster dose estimation using a U-Net model demonstrates high predictive accuracy, with a mean relative deviation of less than 0.4 % per image. Larger deviations are primarily observed in the Bragg peak region and in areas with steep density gradients, as Figure 3 illustrates. Such challenges in regions with abrupt density transitions have also been reported in the context of neural network–based dose calculations [33].

Figure 3 and Figure 4 illustrate that the general shape of the cluster dose distribution is reliably reconstructed. The position of the Bragg peak perpendicular to the beam axis is correctly predicted in most cases, whereas only in a few cases is the predicted peak located in a neighboring voxel. With respect to the depth of the Bragg peak – expressed as the position of the distal 90 % fall-off – maximum deviations of up to nine voxels were observed. These deviations are symmetrically distributed around zero deviation, with a few outliers at lower depths.

In contrast, the height of the Bragg peak is systematically and significantly underestimated, as clearly shown in Figure 4(c). The plateau region is also underestimated, as illustrated by the examples in Figure 3. The models differ in their accuracy in predicting the width of the Bragg peak: while the $g^{F_2}$ model shows only minor deviations of up to about 10 voxels in the FWHM of the Bragg peak, the other models exhibit considerably larger outliers. In all cases, the width is both under- and overestimated to a similar extent.

The computation time for generating the cluster dose distribution with a pencil beam algorithm for a prostate cancer patient was reported to range between 114 s and 1719 s for a carbon ion radiation field [26]. While our reported computation times refer to a single pencil beam, the values from the literature correspond to complete radiation fields, which generally comprise approximately $10^4$ pencil beams. Consequently, the total computation time of the U-Net is in a comparable range to that of the pencil beam algorithm. However, since the U-Net approach is technically straightforward to parallelize, there is potential for substantial reduction in computation time when applied to full radiation fields. A comparison of the algorithm's accuracy is not possible, as the published results are not directly comparable to the data used in this study. However, while the accuracy of pencil beam algorithms is fundamentally constrained by their simplifying assumptions and cannot be further improved, the U-Net approach offers significant potential for improvement, particularly through expanding the training dataset.

Instead of training the U-Net from scratch, this study utilized a pretrained model that had been trained on dose distributions. As numerous studies in this field have shown, such trained networks are increasingly available. Combining these models with transfer learning techniques enables a substantial reduction in the amount of training data required, while also saving computational resources and development time.

Whether this approach can be transferred to other anatomical target regions and clinical scenarios remains the subject of future investigations. Particular attention should be devoted to increasing the variability in the training and test datasets; for example, by incorporating different anatomical regions, beam configurations, and clinical characteristics such as implants or pathological changes. A further point of interest is whether such specific configurations can be captured by cluster dose estimation models when they are already represented in the pretraining dose distributions. Since dose distributions can be generated much more efficiently than cluster dose data, creating a large and diverse dose dataset appears to be a practical and resource-efficient step toward improving the generalizability of such models.

## 5. Conclusion

This proof-of-principle study demonstrates that the computationally intensive MC simulations used for cluster dose calculations can be replaced by a neural network approach. Instead of training from scratch, transfer learning from pretrained dose prediction networks can be employed, significantly reducing the amount of training data that must be generated. The U-Net model used in this work enables cluster dose estimation for a single pencil beam within milliseconds to a few seconds, while maintaining high accuracy. Future studies should investigate the generalizability of this approach to other anatomical regions and treatment modalities, particularly with more extensive and diverse training datasets.

## 6. Acknowledgments

This work was in part supported by the "Metrology for Artificial Intelligence in Medicine (M4AIM)" program, funded by the German Federal Ministry of Economic Affairs and Climate Action in the frame of the QI-Digital Initiative.

## 7. Data Availability Statement

The data that support the findings of this study are openly available at the following URL/DOI: https://gitlab1.ptb.de/MiriamSchwarze/clusterdosetransferlearning.

## Supplementary Figures

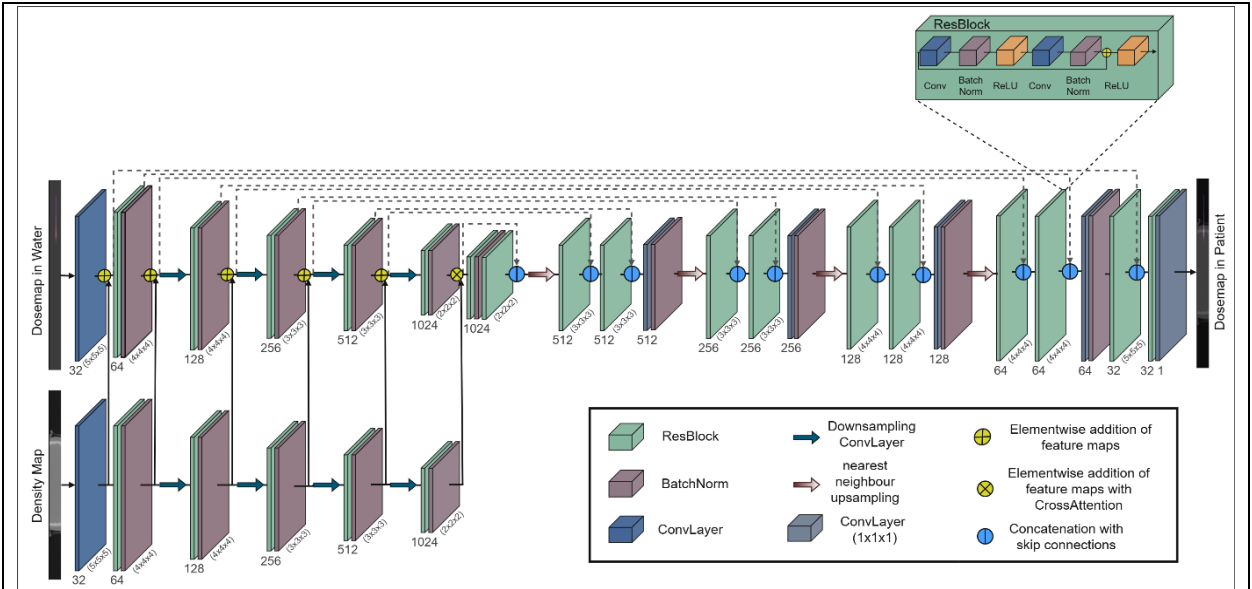

Supplementary Figure 1: Architecture of the dose prediction U-Net used for transfer learning [33].